\documentclass[conference]{IEEEtran}
\usepackage{cite}
\usepackage{amsmath,amssymb,amsfonts}
\usepackage{algorithmic}
\usepackage{graphicx}
\usepackage{textcomp}
\usepackage{xcolor}
\usepackage[hyphens]{url}
\usepackage{multicol}
\usepackage{float}
\usepackage{textcomp}
\usepackage{tikz}
\newcommand\encircle[1]{%
  \tikz[baseline=(X.base)] 
    \node (X) [draw, shape=circle, inner sep=0, fill=black, text=white] {\strut #1};%
}
    
\def\BibTeX{{\rm B\kern-.05em{\sc i\kern-.025em b}\kern-.08em
    T\kern-.1667em\lower.7ex\hbox{E}\kern-.125emX}}

\title{Adaptable Register File Organization for Vector Processors} 

\author{\IEEEauthorblockN{Cristóbal Ramírez Lazo}
\IEEEauthorblockA{\textit{Polytechnic University of Catalonia} \\
\textit{ and Barcelona Supercomputing Center}\\
Barcelona, Spain \\
cristobal.ramirez@bsc.es}
\and
\IEEEauthorblockN{Enrico Reggiani}
\IEEEauthorblockA{\textit{Polytechnic University of Catalonia} \\
\textit{ and Barcelona Supercomputing Center}\\
Barcelona, Spain \\
enrico.reggiani@bsc.es}
\and
\IEEEauthorblockN{Carlos Rojas Morales}
\IEEEauthorblockA{\textit{Barcelona Supercomputing Center} \\
Barcelona, Spain \\
carlos.rojas@bsc.es}
\and
\IEEEauthorblockN{Roger Figueras Bagué}
\IEEEauthorblockA{\textit{Barcelona Supercomputing Center} \\
Barcelona, Spain \\
roger.figueras@bsc.es}
\and
\IEEEauthorblockN{Luis Alfonso Villa Vargas}
\IEEEauthorblockA{\textit{Instituto Politécnico Nacional} \\
Mexico City, Mexico \\
lvilla@ipn.mx}
\and
\IEEEauthorblockN{Marco Antonio Ramírez Salinas}
\IEEEauthorblockA{\textit{Instituto Politécnico Nacional} \\
Mexico City, Mexico \\
mramirezs@ipn.mx}
\and
\IEEEauthorblockN{Mateo Valero Cortés}
\IEEEauthorblockA{\textit{Polytechnic University of Catalonia} \\
\textit{ and Barcelona Supercomputing Center}\\
Barcelona, Spain \\
mateo.valero@bsc.es}
\and
\IEEEauthorblockN{Osman Sabri Ünsal}
\IEEEauthorblockA{\textit{Barcelona Supercomputing Center} \\
Barcelona, Spain \\
osman.unsal@bsc.es}
\and
\IEEEauthorblockN{Adrián Cristal}
\IEEEauthorblockA{\textit{Polytechnic University of Catalonia} \\
\textit{ and Barcelona Supercomputing Center}\\
Barcelona, Spain \\
adrian.cristal@bsc.es}
}

\begin{document}
\maketitle
\thispagestyle{plain}
\pagestyle{plain}


\begin{abstract}

Modern scientific applications are getting more diverse, and the vector lengths in those applications vary widely. Contemporary Vector Processors (VPs) are designed either for short vector lengths, e.g., Fujitsu A64FX with 512-bit ARM SVE vector support, or long vectors, e.g., NEC Aurora Tsubasa with 16Kbits Maximum Vector Length (MVL\footnote{The Maximum Vector Length (MVL) refers to the maximum number of elements held in each vector register. MVL is commonly selected at design time by the computer architect based on the VP target market.}). Unfortunately, both approaches have drawbacks. On the one hand, short vector length VP designs struggle to provide high efficiency for applications featuring long vectors with high Data Level Parallelism (DLP). On the other hand, long vector VP designs waste resources and underutilize the Vector Register File (VRF) when executing low DLP applications with short vector lengths. Therefore, those long vector VP implementations are limited to a specialized subset of applications, where relatively high DLP must be present to achieve excellent performance with high efficiency. To overcome these limitations, we propose an Adaptable Vector Architecture (AVA) that leads to having the best of both worlds. AVA is designed for short vectors (MVL=16 elements) and is thus area and energy-efficient. However, AVA has the functionality to reconfigure the MVL, thereby allowing to exploit the benefits of having a longer vector (up to 128 elements) microarchitecture when abundant DLP is present. We model AVA on the gem5 simulator and evaluate the performance with six applications taken from the RiVEC Benchmark Suite. To obtain area and power consumption metrics, we model AVA on McPAT for 22nm technology.  Our results show that by reconfiguring our small VRF (8KB) plus our novel issue queue scheme, AVA yields a 2X speedup over the default configuration for short vectors. Additionally, AVA shows competitive performance when compared to a long vector VP, while saving 50\% of area. 

\end{abstract}

\section{Introduction}

Supercomputing has always been instrumental as an initial testing ground for innovative architectures. Today, Exascale computing represents  the new milestone for supercomputing. To achieve Exascale performance  within the 20 MW power envelope, highly energy-efficient hardware substrates are needed. VPs are a prime candidate for such substrates as they are typically highly energy-efficient, for example, by computing on operands composed of vectors instead of scalars, therefore requiring fewer instructions to fetch, or by processing multiple vector instructions simultaneously through techniques such as chaining. In that sense, recent Exascale projects have shown a renewed interest in vector architectures.  Some examples are the European Processor Initiative (EPI)~\cite{epi} and the Japanese Post-K~\cite{postk} projects. The EPI project proposes a RISC-V based design, aiming to develop power-efficient and high throughput accelerators. On the other hand, in the Post-K project context, Fujitsu put into operation the Fugaku supercomputer, which is currently ranked first in the TOP500 list ~\cite{top500list}. Fugaku features the Fujitsu ARM A64FX VP, which adopts the ARM Scalable Vector Extension (SVE) ~\cite{arm-sve} as an efficient way to achieve Exascale-class performance.

Although both the ARM SVE and the RISC-V vector extensions took inspiration from the more traditional vector architectures, such as the Cray-1 ~\cite{cray-1}, there is a remarkable difference between them. While ARM SVE allows implementations from 128-bits up to 2048-bits, RISC-V does not limit the MVL, spacing from short and medium size vectors, to long vector designs, which are akin to classic vector supercomputers ~\cite{cray-1,cray-2,cray-x1} and modern VPs ~\cite{BlackWidow,SX-Aurora}. For example, the Aurora VP from NEC ~\cite{SX-Aurora} can multiply-accumulate two 256 element double-precision floating-point vectors in a single instruction.

The vector architectures designed for long vectors are limited to a specialized subset of applications, where relatively high DLP must be present to achieve excellent performance with high efficiency. However, scientific applications are getting more diverse, and the vector lengths in practical applications vary widely. For example, stencil and graph processing kernels usually feature short vectors, while high-performance computing, physics simulation and financial analysis kernels usually operate on long vectors~\cite{tacogem5VA}. We believe that this wide diversity is one of the main reasons behind the trend of building parallel machines with short vectors. Short vector designs are area efficient and are "compatible" with applications having long vectors; however, these short vector architectures are not efficient as longer vector designs when executing high DLP code. 

To help to addressing this wide diversity of vector lengths in practical applications, new vector extensions such as RISC-V V-extension and ARM SVE adopt the Vector Length Agnostic programming. In this programming model, the vector length is not prescribed as in the common Multimedia SIMD ISAs, allowing the vendor to choose the MVL, while guaranteeing portability of the binary code between different hardware implementations. However, since hardware architectures are designed to target specific MVLs, designing for only short or long MVL leads to inefficiencies when trying different DLP patterns. In this paper, we tackle this challenge by proposing a novel vector architecture that combines the area and resource efficiency characterizing short VPs with the ability to handle large DLP applications, as allowed in long vector architectures.

In this context, we present AVA, an Adaptable Vector Architecture designed for short vectors (MVL = 16 elements\footnote{From now on, one element corresponds to a 64-bit word. Thus, the baseline configuration with MVL= 16 elements has a configuration of 1024-bits. The larger configuration with MVL= 128 elements has a configuration of 8192-bits.}), capable of reconfiguring the MVL when executing applications with abundant DLP, achieving a performance comparable to a native\footnote{Native hardware denotes a vector architecture designed for a specific MVL and is the baseline to compare against in this paper. For example, a vector architecture with 64 renamed registers and MVL=128 double precision elements implies a VRF of 64KB.} design for long vectors. The design is based on three complementary concepts. First, a two-level renaming scheme based on a new type of registers termed as Virtual Vector Registers (VVRs), which are an intermediate mapping between the conventional logical and the physical and memory registers. In the first level, logical registers are renamed to VVRs, while in the second level, a VRF-Mapping engine keeps track of which VVRs are mapped to physical registers and which are mapped to memory registers. Second, a two-level VRF, that supports 64 vector registers whose MVL can be configured from 16 to 128 elements. The first level corresponds to the VVRs mapped in the physical registers held in the 8KB Physical Vector Register File (P-VRF), while the second level represents the VVRs mapped in memory registers held in the Memory Vector Register File (M-VRF). While the baseline configuration (MVL=16 elements) holds all the VVRs in the P-VRF, larger MVL configurations holds a subset of the total VVRs in the P-VRF, and maps the remaining part in the M-VRF. Third, we propose a novel two-stage Vector Issue Scheme. In the first stage, the second level of mapping between the VVRs and physical registers is performed, while issuing to execute is managed in the second stage.

Besides the AVA architecture outlined above, the main contributions of this paper are summarized as follows:

\begin{itemize}
\item We demonstrate that AVA improves the performance of applications exploiting both low and high DLP, achieving a speedup of up to 2X by reconfiguring the AVA short-vector implementation.
\item We show that the 8KB P-VRF AVA configuration achieves comparable performance with respect to the equivalent native implementations, which feature a 64KB VRF. 
\item We compare AVA with the Register Grouping (RG) feature proposed by the new RISC-V Vector Extension, and we demonstrate that our scheduling technique can produce fewer swap operations (spill code in RG), performing better in most of the evaluated applications.
\item We show that AVA adds a negligible 0.55\% area overhead, while reducing the total vector processing unit area by 53\% compared with a native design for long vectors.
\item We demonstrate that despite generating additional memory traffic, AVA is energy efficient.
\item We implement the required AVA support at RTL and integrate it on a RISC-V based Vector Processing Unit (VPU), including synthesis and place-and-route at 22nm. 
\end{itemize}

This paper is organized as follows: in Section 2, we present the background and the motivations behind this work. In section 3, we detail our AVA model. In Section 4, the evaluation methodology is shown. The performance, energy and area results are highlighted in Sections 5 and 6, respectively. In Section 7, synthesis and place-and-route experiments are shown. In section 8, related work is described. Finally, Section 9 summarizes the key points of this work. 

\section{Background and Motivation}

An effective way to achieve high performance and efficiency is to leverage on DLP. In this sense, parallel architectures can deliver good performance at a lower cost. One category of parallel hardware organization is termed Single Instruction Multiple Data (SIMD)~\cite{Flynn}. Two variants of SIMD are multimedia extensions and vector architectures~\cite{book-ca}. Multimedia extensions operate on fixed length vector registers. In contrast, in a Vector Architecture, there is no single preferred vector length, just the MVL is defined, and the application can use any vector length that does not exceed the MVL. Nowadays, most commodity CPUs implement architectures that feature SIMD instructions. Common examples for Multimedia extensions include Intel x86’s MMX, SSE, AVX, AVX2 and AVX-512~\cite{intelisa}, MIPS’s MSA~\cite{mipsisa}, ARM’s NEON~\cite{neon}. While classical vector extensions for NEC~\cite{SX-Aurora-guide}  and CRAY~\cite{cray2_isa} are well-known, "the return of the vectors'' includes such contemporary vector architectures as ARM's SVE~\cite{sve}, SVE2~\cite{sve2}, and RISC-V V extension~\cite{riscv-v}.

\subsection{Vector architectures}

Vector architectures, closely identified with supercomputers designed by Seymour Cray~\cite{cray-1,cray-2,cray-x1}, represent an elegant interpretation of SIMD. A key element of these architectures is that arithmetic/logic and load/store instructions operate on sets of vectors instead of individual data items. Moreover, vector architectures typically exploits long execution pipelines to obtain good performance at a lower cost. One of the main features of vector architectures is the VRF, composed of vector registers capable of holding a large number of elements. For these architectures, the maximum number of elements is represented by the MVL parameter, which can vary depending on the hardware implementation~\cite{book-ca}. 

Vector architectures that include multiple lanes can produce multiple results per clock cycle. As shown by Asanović~\cite{asanovic}, adding multiple vector processing lanes is an efficient technique that leads to an advantage in performance and scalability. In a multi-lane vector architecture, each lane synchronously operates with a subset of both the VRF and the functional unit data paths~\cite{vlane-threading}. The VRF usually dominates the area of a single lane, as reported in Ara~\cite{ara} and Hwacha~\cite{embedded0}.

Multi-lane VPs designed for long vectors achieve excellent computational throughput and the most efficient execution for programs with high DLP. However, applications lacking abundant DLP are unable to fully utilize the hardware resources in the vector lanes. When the Application Vector Length  is notably smaller than the MVL, multiple inefficiencies arise. First, short vector applications cannot fully use each vector register width, as a portion of each vector register remains underutilized during the whole program execution. Second, when the number of vector registers is not sufficient, the compiler generates spill code.  At compilation time, the compiler is not aware of the Application Vector Length. In that sense, the spill code includes load/store of vector registers with the MVL, even though the application only needs a portion of them. This behavior could lead to a performance degradation as well as energy waste. 

Long vectors bring several advantages such as maximizing the amount of latency amortized per vector instruction. In that sense, different ideas have been studied trying to preserve multi-lane VPs designed for long vectors, while being able to exploit different DLP patterns in an efficient way by reconfiguring the available resources. A couple of the more representative examples for this related work are described below.

Krashinsky et al. proposed the Vector Thread Architecture~\cite{vthread-arch}, a hybrid multithreaded vector architecture that provides a control processor and an array of slave virtual processors to the programmer. To execute high DLP code, the control processor can use vector-fetch commands to broadcast vector instructions to be executed in all virtual processors, where each virtual processor executes a subset of the vector elements as in the traditional multi-lane designs. On the contrary, to execute thread-parallel code, each virtual processor can use thread-fetches to direct its own control flow as an efficient way to execute short vectors.

Rivoire et al. proposed Vector Lane Threading ~\cite{vlane-threading}, an architectural enhancement that allows idle vector lanes to run short-vector or scalar threads. When running low DLP code, they assign the different lanes across several threads. Then, the combination of threads can saturate the available computational resources. In that sense, the microarchitecture allows the exploitation of data-level and thread-level parallelism to achieve higher performance. 

While the above approaches also feature some reconfigurability, their base VP design targets long vectors, which is costly in terms of area and resources. In contrast, AVA is centered around a design targeting short vectors, which is inherently area and resource-efficient. However, AVA reconfigurability enables this short vector design to perform as well as a VP designed for long vectors.  Additionally, featuring a small VRF offers several advantages, such as the opportunity to implement multi-ported memory structures, a feature that for large memory structures could be highly costly in terms of area and power, or sometimes prohibited depending on the design requirements.

\begin{figure*}[h!]
\centering
\includegraphics[width=1.0 \textwidth]{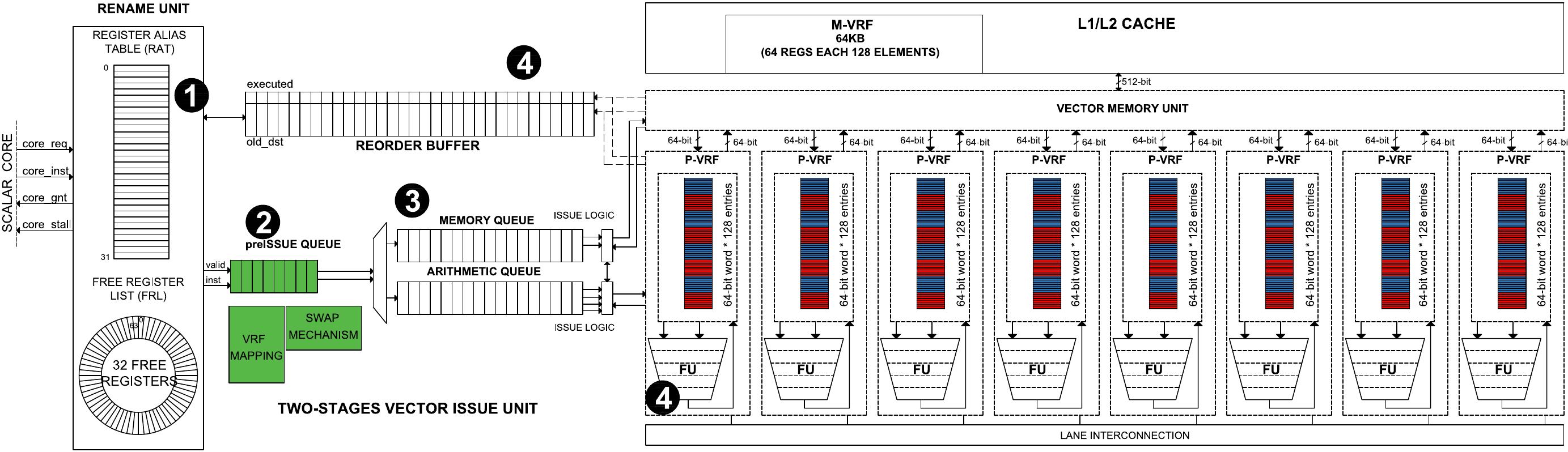}
\caption{AVA microarchitecture overview. The new hardware additions are highlighted in green, involving the second stage of the renaming unit (VRF-Mapping) and the first stage of the Vector Issue Scheme (pre-issue queue and Swap Mechanism).}
\label{figure1}
\end{figure*}

MVL reconfigurability has also been proposed at ISA level. For example, the new RISC-V vector extension~\cite{riscv-v} includes a novel feature called Register Grouping (RG), whose main goal is to increase the execution efficiency for applications featuring high DLP. RG allows grouping multiple vector registers together, so that a single vector instruction can operate on multiple vector registers as if it was a single "wider" register at the cost of having fewer available architectural registers. The Vector Length Multiplier (LMUL) represents the default number of vector registers that are combined to form a vector register group. Specifically, LMUL supports four different configurations (i.e., 1,2,4,8). For those values, the MVL can be increased by 1x, 2x, 4x and 8x while reducing the number of architectural registers from 32 defined by the vector ISA down to 16, 8 and 4, respectively. It is worth noticing that when the application needs more architectural registers than the one available at that time, spill code is generated by the compiler. The bigger the LMUL configuration, the higher the probability  of generating spill code. When implementing renaming, physical vector registers are also reduced by LMUL. This implies that for a renaming of 64 physical vector registers for the LMUL=8 configuration there are only 8 register groups available, 4 assigned initially in the Register Alias Table (RAT) and 4 in the Free Register List (FRL). This leads to accepting only four vector instructions before the FRL is empty, and a stall occurs in the scalar core. 

Similarly, AVA pursuits the same RG goal, which is to provide the capability to operate on longer vectors when applications exhibit abundant DLP. However, AVA allows this reconfigurability completely at hardware level, preserving the architectural registers (32 logical vector registers) regardless of the MVL configuration. Therefore, AVA can accept as many vector instructions as the number of free registers it has. Larger instruction windows give the opportunity to exploit ILP. Additionally, while RG is an exclusive feature of RISC-V, AVA can be implemented over different microarchitectures regardless the target vector ISA. 

Vector architectures have been proposed for embedded systems~\cite{embedded0,manic,embedded1} despite their popular association with high-area. In fact, VPs are also suitable and even more efficient for power-constrained embedded systems, since vector execution provides energy-efficiency benefits of amortizing instruction supply energy (fetch, decode, and issue) across many operations. Furthermore, even though a larger VRF incurs higher access energy, longer VLs are still beneficial for embedded applications, as established by Gobieski et al.~\cite{manic}. AVA perfectly matches with embedded systems, since one of the main ideas is to implement a small VRF, while able continue executing long vectors.

Maximizing the use of expensive hardware resources such as vector registers is an important goal in the computer architecture community, since energy efficient hardware is required from the HPC market to achieve Exascale levels to the embedded market for ultra-low-power embedded systems. 

\section{Adaptable Vector Architecture (AVA)} 

As Fig.~\ref{figure1} shows, the default configuration of the AVA microarchitecture supports 64 physical vector registers, having a MVL of 16 elements, and thus composing the P-VRF of 8KB distributed between 8 lanes. However, the adaptability of the proposed architecture allows to scale the MVL ranging from 16 to 128 elements, while consuming the same P-VRF size and the same number of VVRs. To enable this feature, the P-VRF is complemented by a M-VRF that does not have direct access to the functional units. Specifically, when the MVL is equal to 16 elements, all the 64 VVRs are held in the P-VRF and none in the M-VRF. Instead, when the MVL is greater than 16 elements, the VVRs are distributed among the P-VRF and the M-VRF. For example, when the MVL is equal to 128, 8 VVRs are held in the P-VRF, while the remaining 56 are allocated in the M-VRF. The interaction between the P-VRF and the M-VRF is handled by the following components: (1) a two-level renaming scheme, composed of a first level that maps the 32 logical registers (ISA registers) to the 64 VVRs, and by a second level, that maps the VVRs to the physical registers located in the P-VRF and/or to the memory registers located in the M-VRF; (2) a two-level vector issue scheme, the first level of which determines which, if any, VVR of the issuing instructions need to be swapped-in from the M-VRF to the P-VRF, while the second level manages the issuing to execution.

As a general example of how AVA modules interact, Figure 1 shows the life cycle of one vector instruction in AVA modules, denoted by steps from 1 to 4. In \encircle{1} the instruction arrives to the renaming stage, where the instruction operands are renamed from logical registers to VVR.  In the next stage, the instruction payload is sent to the pres-issue stage. At \encircle{2} the pre-issue stage is in charge of mapping the VVRs to physical registers. If the source operands are located in the M-VRF, a Swap Mechanism moves the related VVRs from the M-VRF to the P-VRF. Additionally, one physical register is assigned in case the vector instruction requires one physical register to write-back the result. In case there are no available physical registers, the Swap Mechanism selects and copies one VVR located in the P-VRF to the M-VRF, thus freeing a physical register for the instruction. Once the vector instruction operands have been renamed to physical registers, the vector instruction is sent to the second issue stage  \encircle{3}, which consists of the arithmetic and memory queues. \encircle{4} Once the instruction is issued and executed, the Reorder Buffer marks it as executed, and waits for its turn to commit.

The next subsections describe the three key components of AVA in more detail (two-level renaming scheme, two-level vector register file and two-stage vector issue unit), followed by a detailed functional description of the overall design.

\subsection{Two-level Renaming Scheme: Virtual, Physical and Memory Registers} 

AVA implements a two-level renaming scheme which is based on a new type of registers termed as VVRs, which are an intermediate mapping between the logical registgers and the physical and memory registers.

In the first level, logical registers are renamed to VVRs using the conventional structures: The RAT, a 6-bit x 32-entries structure in charge of keeping the mapping between the logical registers and the VVRs, and the FRL which contains the available VVRs to be assigned as a virtual destination.

\textit{ Freeing up Virtual Vector Registers}. Freeing up VVRs is performed when an instruction commits. Then, the corresponding old destination VVR is sent to the FRL. Additionally, the corresponding Register Access Counter (RAC) (see section III.C for RAC details ) is set to 0. 

In the second level, the VRF-Mapping logic keeps track of which VVRs are either mapped to physical or memory registers. This logic is composed of three simple structures. The First structure is the Physical Register Mapping Table (PRMT), a 6-bit x 64-entries structure in charge of keeping the correspondences between the VVRs and the physical registers. The second structure is the Vector Register Location Table (VRLT), a 1-bit x 64-entries structure that indicates if a given VVR is located in the physical or memory registers. Third, the Physical Free Register List (PFRL) is a structure that holds the available physical registers to be assigned. 

\textit{ Freeing up Physical Registers}. The freeing up of a physical register occurs in two distinct cases. (1) AVA exploits the concept of aggressive register reclamation~\cite{Akkary} to enable physical register usage to closely match the true lifetime of registers. In this sense, it is possible to claim and free a physical register that will not be longer used. The aggressive register reclamation is applied only when: (a) a RAC (see section III.C for RAC details) counter reaches zero, meaning that a specific VVR has been renamed, that all the consumers have read the VVR, and that the VVR has become an old destination of a younger instruction, and  (b) there are no older vector memory instructions in the pipeline. In this scenario, the corresponding physical register assigned to the VVR which has its count equal to zero can be pushed to the PFRL structure. Note that by updating the RAC counters at commit time we ensure that the freeing up will not create a conflict in case a recovery event (branch missprediction or exception in the scalar pipeline) arises. This is because we are ensuring that all the instructions that read that VVR has committed. (2) When a physical register for the new instruction is needed, but there is no RAC count with 0. In this case, based on the information provided by the RAC counters, it is selected the VVR mapped in the P-VRF which has the lower count, and does not matches with any of the instruction virtual source operands, to be sent to the M-VRF and freed the corresponding physical register. 

Contrary to the RISC-V RG proposal where the number of logical and physical registers are reduced by LMUL factor, our model allows to preserve the same number of Logical and VVRs no matter if the MVL increases.

\subsection{Two-level Vector Register File} 

The adaptability of AVA allows to reconfigure the MVL from 16 elements up to 128 elements while keeping the same modest P-VRF size. It is achieved by backing the P-VRF with a second level VRF termed as M-VRF. In this scheme, the VVRs that are being used or close to be used are assigned to the first level (i.e., P-VRF) allowing them to have direct access to the functional units. On the other hand, the VVRs that are not being used or will not be used soon are assigned to the second level (i.e., M-VRF). Additionally, each VVR is associated with one entry of the valid-bit structure (1-bit x 64-entries) which indicates a valid data. When a VVR is assigned at renaming time, the associated Valid-bit is set to 0. Once the vector instruction executes, the associated Valid-bit is set to 1.

Since our baseline microarchitecture features an 8-Lane VPU, the P-VRF is implemented as eight 4R-2W 1-KB (64-bit words x 128 entries) SRAM memory structures distributed between the eight lanes. The P-VRF contains 64 physical registers where each register is 16 elements wide for the baseline configuration as illustrated in Figure 1. Note that our model is restricted to execute one arithmetic operation plus one memory operation in parallel. Accordingly, 3 read ports and 1 write port are assigned to the arithmetic pipeline, while 1 read port and 1 write port are assigned to the memory unit. Adding more arithmetic pipelines would increase the required VRF ports which has a super-linear impact on the power/area results as demonstrated by Arima et al.~\cite{arima} and Zyuban et al~\cite{rf-complex}. 

\begin{scriptsize}
\begin{table}[h!]
  \centering
  \caption{Physical Vector Register File Configurations.}
  \label{table:table1}
  \begin{tabular}{ | c | c | c | c | c | c | c | c | c | }
    \hline
    P-Regs & 64 & 32 & 21 & 16 & 12 & 10 & 9 & 8 \\
    \hline
    MVL & 16 & 32 & 48 & 64 & 80 & 96 & 112 & 128 \\
    \hline
  \end{tabular}
\end{table}
\end{scriptsize}

Furthermore, by setting a configuration register, it is possible to configure the VPU for longer MVLs at the cost of reducing the number of physical registers that can be held in the P-VRF. For example, we can configure from 64 physical registers (16 elements each), down to 8 physical registers (128 elements each) in multiples of 16 elements as summarized in Table~\ref{table:table1}. Note that supporting all the proposed configurations does not incur in additional routing overhead. Indeed, the read/write control logic iterates MVL/lanes times until it completes the read/write operation. 

When MVL$>$16 elements, it is needed to reserve the required memory for holding the M-VRF. In our experiments, we use a custom intrinsic called set\_virtual\_vrf  (performing a malloc assignment and sending the base address to the VPU) to reserve memory for the M-VRF. However, ideally, the OS takes care of reserving the memory space for each thread.

\subsection{Two-stages Vector Issue Unit } 

The two-stages vector issue unit is composed of the pre-issue stage and the issue stage.  We now explain both stages in turn. 

Pre-issue stage: The first level of mapping from Logical Registers to VVF occurs in the renaming stage. Pre-issue stage performs the second level of mapping between the VVRs and physical registers.
As mentioned before, when MVL$>$16, a subset of the VVRs is held in the P-VRF, while the remaining VVRs are allocated in the M-VRF. In case a new physical register is required, but there is not any free physical register, the content of a selected VVR is sent to the M-VRF to free one physical register, and is assigned to the new instruction. Eventually, VVRs previously moved to the M-VRF can be needed by a new vector instruction, which then requires to move the content back to the P-VRF. We term these operations as Swap operations. In consequence, AVA implements a SWAP Mechanism module which is composed of two main structures : the RAC and the Swap Logic. (1) The RAC is a 3-bit x 64-entry structure where each entry holds how many times a specific VVR is read. At the first stage of the renaming, the RAC counters are updated. First, the new destination and source VVRs increment the corresponding register count, while the old destination VVR decrements the corresponding count. At commit time, the counters are updated again. This time, the source VVRs decrement the corresponding counter. The RAC helps to take decisions based on the count of each individual VRR which are described in the next paragraphs. (2) The Swap Logic decides which VVRs should be swapped to the M-VRF, and creates memory operations termed as Swap-Stores. Swap Logic also decides when it is required to move VVRs from the M-VRF to the P-VRF, and creates operations termed as Swap-Loads.  The Swap Logic takes advantage of the information provided by the RAC counters to decide which VVR allocated in the P-VRF should be swapped to the M-VRF. This selection is based on the VVR mapped in the P-VRF which has the lower count (1 is the lowest count for swaps, and 0 is the count for aggressive register reclamation), and selection logic also checks that the candidate VVR does not match with any of the instruction virtual source operands to avoid deadlock. 

Pre-issue stage implements an in-order issue scheme.  A vector instruction is ready to be issued to the second level only when it has been fully renamed from VVRs to physical registers. However, renaming the instruction from VVRs to physical registers implies several steps  evaluated in the following order: 

\begin{figure*}[h!]
\centering
\includegraphics[width=0.7\textwidth]{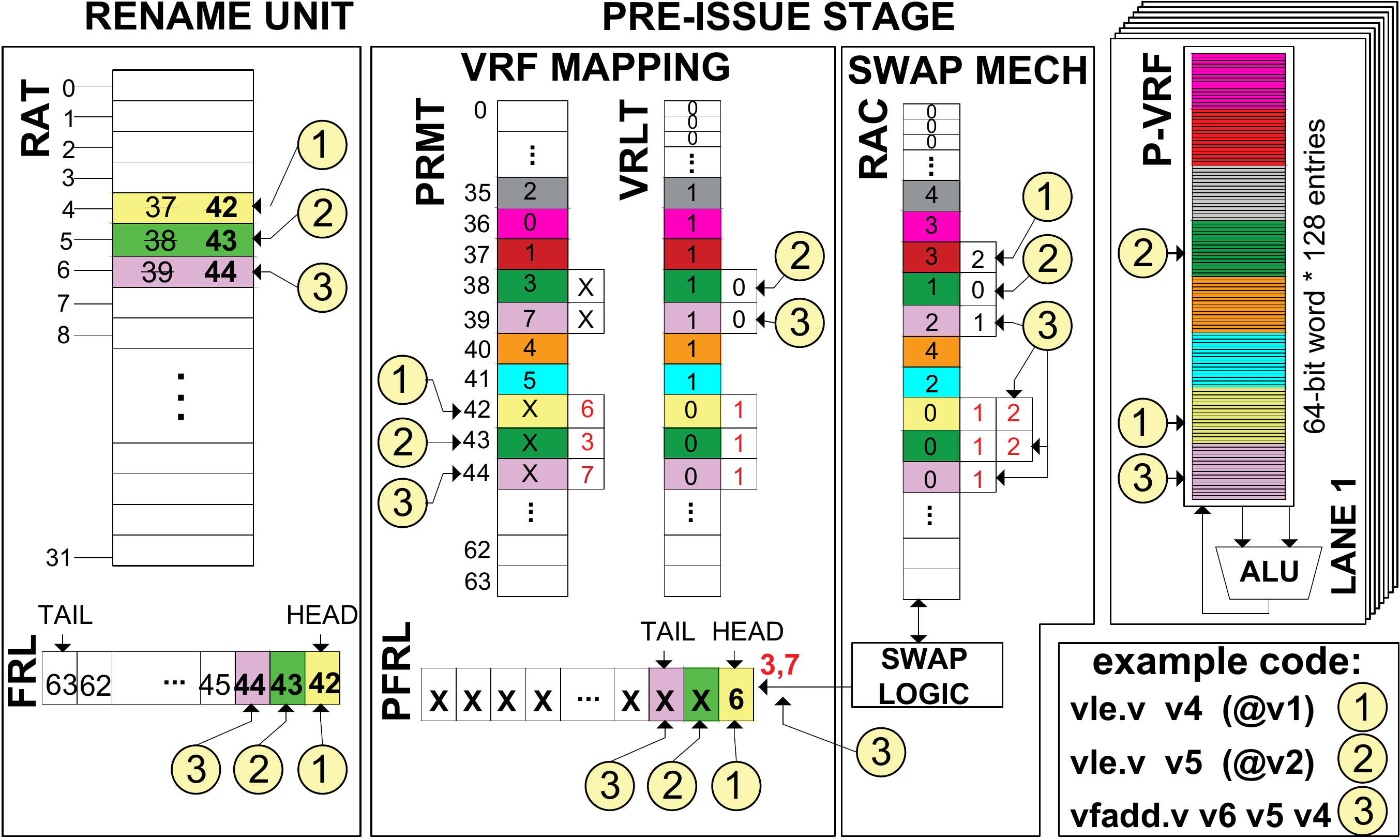}
\caption{Register Mapping example}
\label{figure2}
\end{figure*}

(A) Source VVRs are mapped to the corresponding physical register by reading the PRMT and the VRLT structures (indexed by the source VVRs). There are two possible scenarios for each source operand: (1) If the value read from the VRLT is equal to "1", indicates that the physical register obtained from the PRMT structure is valid and it is located in the P-VRF, and the corresponding source VVR can be mapped immediately. (2) On the  contrary,  if the value read from the VRLT is equal to "0", the VVR is located in the M-VRF and is loaded to the P-VRF to be used. In this second scenario, it is required to notify the conflict to the Swap Mechanism.  A couple of tasks are performed by the Swap Mechanism: (Swap-1)  Verifies that there is at least one physical register available to load the values from the M-VRF. In case there are not free physical registers,  a Swap-Store is created and sent to the Memory queue to store the content of one VVR selected by  the Swap Logic from the P-VRF to the M-VRF.  With this, the associated physical register can be freed and pushed to the PFRL. (Swap-2) Then, a Swap-Load is created and sent to the Memory queue to load the VVR from the M-VRF to the P-VRF. 

(B) If the vector instruction requires to write-back its result, a new physical register must be assigned. In case there are no free physical registers, the task Swap-1 must be repeated. Then, the new available physical register can be assigned as the physical destination.

(C) Finally, once the instruction has been renamed, it is issued to the second level only if there is availability in their corresponding queue. Otherwise, a stall is signaled until there is at least one free slot for the instruction.

Issue stage: it is composed of the Memory and Arithmetic Queues in charge of issuing the vector instruction. Individually each queue performs in-order issue, however, since the memory queue is decoupled from the Arithmetic queue, there is a light out-of-order behaviour. Because of the introduction of Swap Operations, AVA must guarantee for each instruction either in the Arithmetic or Memory Queue to have its source VVRs mapped to the P-VRF when it is its turn to be issued to execute, avoiding deadlocks. This is done by following 2 rules:  
(1) Swap-Stores created to free one physical register must notify to the new owner of the physical register that it has executed, meaning that the VVR previously mapped in the physical register is now in the M-VRF, and then it is possible to write-back new data to the physical register. (2) Swap-Loads must wait to all the consumers of the previously VVR mapped in P-VRF have read the register.

Once the instruction is issued and executed, it will be marked in the reorder-buffer as executed, only waiting for its turn to commit.

\subsection{Recovering the microarchitectural state} 

After some event such as a miss-prediction or memory exceptions, AVA can roll back and recover the microarchitecture state. The renaming tables (RAT and FRL pointers) and the Valid-bit are the only mandatory structures to be recovered. Therefore, AVA implements only one copy that is updated every time a vector instruction commits.

Recovering the RAC counters is optional, since every time that a VVR is freed, the respective count is also set to zero. Thus, not recovering the state of the counters does not imply any correctness issue.

\subsection{AVA Functional Description} 

Fig.~\ref{figure2} illustrates the AVA functional behavior based on the execution of three instructions.  The selected MVL configuration is 128 elements, meaning that only 8 physical registers are available. Also, to exemplify how the Swap Mechanism works, we assume that several vector instructions were executed previously, so that some physical registers were previously assigned to older instructions.

Once the scalar core sends the first instruction to the decoupled VPU, it is received by the first stage of the renaming unit. In this stage, the logical registers are renamed to VVRs. Since the instruction is a vector load, only the destination logical register 4 reads the RAT to obtain the associated old destination VVR 37. The  destination VVR 42 is obtained from the FRL. In parallel, the new destination VVR 42 increments the corresponding RAC entry, while the old destination VVR 37 decrements the corresponding RAC entry.  In the next cycle, the instruction advances to the pre-issue stage. Since it is a load, the only requirement is to obtain a physical register to be used as a destination. At this moment, the PFRL has the physical register 6 available, which is assigned as the physical destination. Then, this new mapping is written in the location 42 of the PRMT. Additionally, the corresponding entry in the VRLT is set to 1, indicating that the VVR 42 is now mapped in the physical registers. After this, the instruction is sent to the Memory Queue in second stage to wait for execution.

As also the second instruction is a vector load, performing the same process in the renaming unit as the previous load, the VVR 43 is assigned as the virtual destination. In parallel, the new destination VVR 43 increments the corresponding RAC entry, while the old destination VVR 38 decrements the corresponding RAC entry. Note that after the subtraction the count reaches 0, meaning that it is possible to reclaim physical register 38 to be used for a new physical destination, since it is guaranteed that the value corresponding to the VVR 38 will be never used by any later instruction. Additionally, the location 38 in the VRLT is set to 0 indicating that the VVR 38 is no longer mapped in the physical registers. In the following cycle, the instruction advances to the pre-issue stage, where the PFRL points to physical register 3 as being available, which is assigned as a physical destination for the load. Then, the instruction is sent to the Memory Queue in second stage waiting to be executed.

The last instruction corresponds to a vector addition. This time, the sources and destination logical registers read the RAT to obtain the associated source VVRs 42 and 43, and old destination VVR 39 respectively.  The VVR 44 is assigned as the destination. In parallel, the source VVRs 42 and 43, and the new destination VVR 44 increments the corresponding RAC entry, while the old destination 39 decrements the corresponding RAC entry. After that the subtraction the count reaches 1, which means that this time it is not possible to reclaim physical register 39. In the next cycle, the instruction advances to the pre-issue stage, where now the PFRL does not have any physical register available, and any counter of the VVRs mapped in the physical registers has reached 0. Then, this forces a swap operation. To do that, the RAC entry with the smaller count is selected, which corresponds to the VVR 39 as the register that will be sent to the memory registers. Subsequently, a Swap-Store operation is created and issued to the memory queue to send the content of the VVR 39 to the memory registers. Finally, the physical register 7 is freed and pushed to the PFRL, to be assigned later as a physical destination for the vector addition. Then, the instruction is sent to the Arithmetic Queue in second stage and wait for being executed. Once every instruction commits, all source VVRs will decrease the associated RAC entry by one.

\section{Evaluation Methodology}

\begin{scriptsize}
\begin{table*}[h!]
  \caption{System configurations}
  \label{table:table2}
  \centering
  \begin{tabular}{ | c | c | c | c | c |}
    \hline
    \textbf{NATIVE X1} & \textbf{NATIVE X2} & \textbf{NATIVE X3} & \textbf{NATIVE X4} & \textbf{NATIVE X8} \\
    \hline
    \hline
    \multicolumn{5}{|c|}{\text{Dual-Issue 64-bit RISC-V superscalar in-order pipeline, Clock Frequency - 2 GHz}}\\
    \hline
    \multicolumn{5}{|c|}{\text{VPU with 8 Lanes (1 pipelined arithmetic unit/Lane)  - Clock Frequency  - 1 GHz}}\\
    \hline
    \multicolumn{5}{|c|}{\text{8 Lanes (1 pipelined arithmetic unit / Lane)}}\\
    \hline
    MVL 1024-bit & MVL 2048-bit & MVL 3072-bit & MVL 4096-bit & MVL 8192-bit \\
    (16 elem * 64-bit) & (32 elem * 64-bit)  & (48 elem * 64-bit) & (64 elem * 64-bit) & (128 elem * 64-bit)\\
    \hline
    \multicolumn{5}{|c|}{\text{64 Renamed Registers}}\\
    \hline
    4R/2W VRF: 8KB & 4R/2W VRF: 16KB & 4R/2W VRF: 24KB & 4R/2W VRF: 32KB & 4R/2W VRF: 64KB \\
    \hline
    \multicolumn{5}{|c|}{\text{Vector Arithmetic and Memory Queue - 32 entries each}}\\
    \hline
    \multicolumn{5}{|c|}{\text{VMU connected to L2 Bus, 512-bit memory interface}}\\
    \hline
    \multicolumn{5}{|c|}{\text{Memory System}}\\
    \multicolumn{5}{|c|}{\text{32KB L1I , 32KB L1D – Latency 4 cycles – cache line 512-bit}}\\
    \multicolumn{5}{|c|}{\text{1MB L2 – Latency 12 cycles – cache line 512-bit, 2 GB DDR3 Memory}}\\
    \hline
  \end{tabular}
\end{table*}
\end{scriptsize}

\begin{scriptsize}
\begin{table*}[h]
    \caption{AVA and RG configurations  and their corresponding equivalence with the five configurations in Table II.}
  \label{table:table3}
  \centering
  \begin{tabular}{ | c | c | c | c | c |}
    \hline
    \textbf{NATIVE X1} & \textbf{NATIVE X2} & \textbf{NATIVE X3} & \textbf{NATIVE X4} & \textbf{NATIVE X8} \\
    \hline
    \hline
    AVA X1 (64-PREG) & AVA X2 (32-PREG) & AVA X3 (21-PREG) & AVA X4 (16-PREG) & AVA X8 (8-PREG)\\
    \hline
    RG-LMUL1 & RG-LMUL2 & NA & RG-LMUL4 & RG-LMUL8\\
    \hline
  \end{tabular}
\end{table*}
\end{scriptsize}

\begin{scriptsize}
\begin{table}[h!]
  \caption{Selected applications from RiVEC Benchmark Suite}
  \label{table:table4}
  \centering
  \begin{tabular}{ | c | c | c | }
    \hline
    \textbf{Application} & \textbf{Application} & \textbf{Algorithmical}\\
                         & \textbf{Domain}      & \textbf{Model}\\
    \hline
    \hline
    Axpy & HPC & BLAS\\
    \hline
    Blackscholes & Financial Analysis & Dense Linear Algebra\\
    \hline
    LavaMD2 & Molecular Dynamics & N-Body\\
    \hline
    Particle Filter & Medical Imaging & Structured Grids\\
    \hline
    Somier & Physics Simulation & Dense Linear Algebra\\
    \hline
    Swaptions & Financial Analysis & MapReduce\\
    \hline
  \end{tabular}

\end{table}
\end{scriptsize}

To evaluate our ideas, we use as a base platform a parameterizable decoupled vector architecture~\cite{decoupled} based on the RISC-V Vector extension modeled on the gem5 simulator~\cite{gem5-vector,tacogem5VA}. We added the necessary modifications to implement the AVA architecture, substantially modifying the issue stage which also includes the queues, the swap mechanism, and the VRF read/write logic.

Table~\ref{table:table2} presents five system configurations where a VPU is attached to a scalar core. The VPU configurations vary the MVL's. NATIVE X1 corresponds to the baseline hardware with 64 physical registers with MVL=16 elements (1024-bits), leading to a VRF size of 8KB. The remaining configurations (from NATIVE X2 to NATIVE X8) represent a costly hardware implementation, increasing the MVL size in every configuration leading to VRF sizes from 16-KB up to 64-KB for the larger configuration.

Table~\ref{table:table3} shows five different AVA configurations. AVA X1 represents the baseline model (64 physical register with MVL=16 elements). AVA X2 to AVA X8 represents the AVA configurations that after reconfiguring AVA X1 match the NATIVE X2 to NATIVE X8 configurations. Also, the number of physical registers available for each configuration is shown. In the same way, the equivalent configurations for RISC-V RG are listed (LMUL1, LMUL2, LMUL4, and LMUL8). It is important to emphasize that both AVA and RG proposals use the baseline configuration with an 8-KB VRF for all their configurations. 

We choose the RiVEC Benchmark Suite~\cite{bsuite,tacogem5VA} to evaluate our AVA proposal since it is a benchmark suite originally designed to evaluate vector architectures and covers a wide spectrum of domains as shown in Table~\ref{table:table4}. In this suite, all the applications are hand-vectorized by using RISC-V vector intrinsics. For all the applications, we have compiled four versions. The first version is compiled using the flag for LMUL=1. The resulting binary is used to evaluate the baseline configuration (MVL=16), and all the AVA and NATIVE configurations. The following ones use the flags to compile for the LMUL2, LMUL4, and LMUL8 configurations respectively.

To obtain the main physical metrics such as area and energy we have extended the McPAT framework to model AVA. We also model all the configurations shown in Table 3. Finally, AVA was successfully implemented at RTL level on the Hydra VPU, more details are shown in Section 7.

\section{Performance Evaluation}

\begin{figure*}[h!]
\centering
\includegraphics[width=1.0\textwidth]{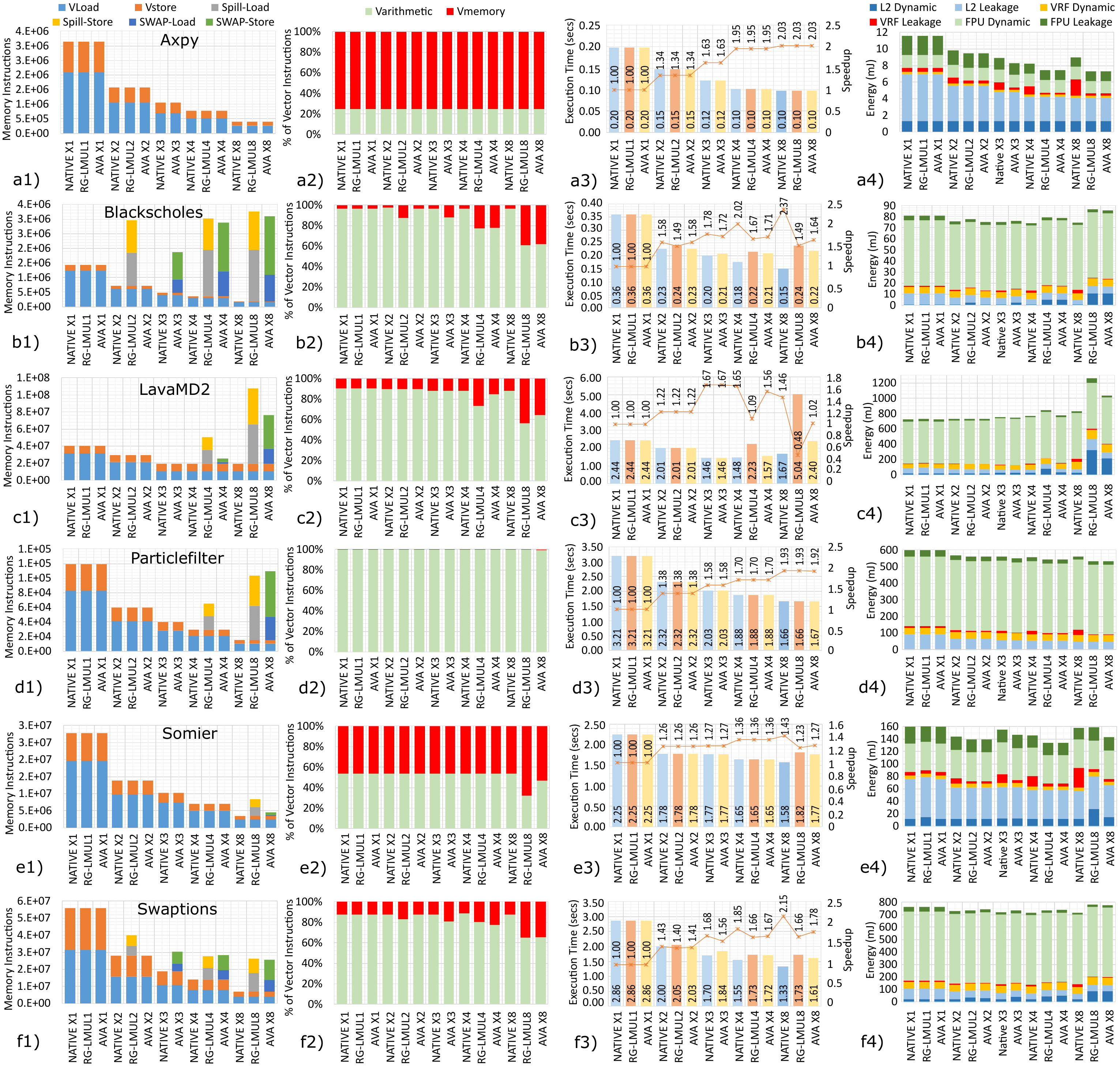}
\caption{Performance evaluation for the six applications: a)Axpy; b)Blackscholes; c) LavaMD2; d) Particle-Filter; e)Somier; f)Swaptions.  Charts in the first column shows the Vector Memory Instruction count including SPILL operations generated by the compiler and SWAP operations generated by AVA. Charts on the second column shows the \% of vector instruction, charts on the third column shows the Execution-time and Speedup when comparing versus NATIVE-1 (baseline), and charts on the fourth column shows the Energy consumption obtained from McPAT.}
\label{figure3}
\end{figure*}

The first application is Axpy, which represents the ideal scenario for both RG and AVA where RG-LMUL8 and AVA X8 obtain the same performance compared to a native VPU designed for long vectors (NATIVE X8) , and achieving a speedup of 2X with respect to the baseline configuration (NATIVE 1), as illustrated in Figure~\ref{figure3}-a3. Also, as shown in Figure~\ref{figure3}-a1, neither spill code from the compiler nor swap operations from the Swap-Logic are created since Axpy only uses two logical vector registers. Figure~\ref{figure3}-a2 shows the percentage of vector arithmetic and memory instructions. For all the configurations, the vector memory instructions (Vmemory) represent 75\%, and the vector arithmetic instructions (Varithmetic) represent 25\% of the total vector instructions.

The second application is Blackscholes. This high DLP application is interesting to analyze since the vector compiler makes use of 23 logical vector registers to obtain the final binary. At first glance, we can see that there is high pressure in the use of vector logical registers. For LMUL=2,4 and 8, the compiler can make use of only 16,8 and 4 logical vector registers respectively, and for any of those configurations, spill code is added as shown in Figure 3-b1. AVA presents a similar behavior. However, it is interesting to see that for AVA X2 there are no swap operations. This is because the scheduling is done using 32 physical vector registers, meaning that we have enough vector registers to compute the application without generating swap operations. On the other hand, swap operations are generated starting from the AVA X4. Also, the number of swap operations is slightly less than the number of spill code operations generated by the compiler. This is because AVA performs the scheduling based on the available physical registers, which are always double compared to LMUL. Figure 3-b3 shows the performance results. For AVA-X2 there are not swap operations, thus a similar performance to NATIVE-2 is achieved, and a speedup of 1.58X over the baseline configuration.  AVA X4 achieves a speedup of 1.71X over the baseline configuration.  For AVA X8, the percentage of memory operations represents 38\% of the total vector instructions, causing a slight performance degradation (1.64X) because of the increased number of swap operations. For all the configurations AVA performs better than RG since less memory traffic is generated.

For LavaMD2, the vector compiler uses 15 logical vector registers to create the final binary, which implies that for RG-LMUL2, no spill code is necessary. However, for RG-LMUL4 and RG-LMUL8, spill code is generated as shown in Figure 3-c1, causing an increase in memory operations from 9\% for RG-LMUL1 configuration, to up to 43\% for RG-LMUL8 configuration as shown in Figure 3-c2. For AVA, the SWAPS operations are few compared with the equivalent spill code generated by the RG-LMUL configuration. Figure 3-c3 shows the performance results. First, this application makes use of a fixed vector size of 48 elements, meaning that for the configurations with a larger MVL than 48 elements we cannot make full use of each vector register, and a portion of each vector register remains unused during all the program execution. For AVA the best configuration is AVA X3. AVA X3 not only executes the 48 elements with only one instruction, but also 21  physical registers are available for the computation, thereby avoiding swap operations, as shown in Figure 3-c1. Also, it achieves a speedup of 1.67X, better than any of the RG- LMUL configurations and equal to the equivalent NATIVE configuration, as shown in Figure 3-c3. Finally, another interesting result is for RG-LMUL8, where the performance decrease notably. The reason is because for this configuration, the memory operations represent 43\% of the overall vector instructions. Also, 81\% of the memory operations are spill code. As described in Section 3.1, the spill code is always executed using the MVL. As a result, the memory operations become the bottleneck since all the arithmetic operations (57\%) are executed with VL=48, while spill code is executed with VL=MVL=128. 

For Particle-Filter, the compiler requires 13 logical vector registers to generate the final binary, which implies that for RG-LMUL2, AVA X2, and AVA X3, no spill/swap operations are added. On the other hand, spill/swap operations are generated for RG-LMUL4, RG-LMUL8, AVA X4, and AVA X8, as shown in Figure 3-d1. However, the increase in memory operations percentage is negligible, representing 0.15\% for the larger configuration, achieving similar performance levels as the corresponding NATIVE configuration as shown in Figure 3-d3. 

For Somier, the vector compiler uses 13 logical vector registers to generate the final binary. Spill/swap operations are generated only for RG-LMUL8 and AVA X8. For RG-LMUL8 there was an increase in the percentage of memory operations from 46\% to 68\% as shown in Figure 3-e2. For AVA X8, few swap operations were generated. Figure 3-e3 shows the performance results. In this case, the NATIVE X4, RG-LMUL4, and AVA X4 achieves the best speedup with 1.43X. For AVA X8 and RG-LMUL8 there was a small performance degradation because of the additional memory traffic.

Finally, for Swaptions, the vector compiler uses 24 logical vector registers to generate the final binary, which implies that for RG-LMUL2, RG-LMUL4, and RG-LMUL8 spill code is generated as shown in Figure 3-f1, causing an increase in the percentage of memory operations from 12\% in the NATIVE-1 configuration up to 34\% in the RG-LMUL8 configuration as shown in Figure 3-f2. For AVA, the swap operations appear starting from AVA X3, obtaining almost the same number as the compiler generated spill code for RG. AVA-8 achieves a speedup of 1.78X while the NATIVE-8 configuration achieves 2.15X with respect to the NATIVE-1 configuration.  

As shown above, AVA provides performance improvements for all the evaluated applications, being competitive with NATIVE designs for longer vectors. 

\section{McPAT Area and Energy Evaluation}

\begin{figure}[h!]
\centering
\includegraphics[width=0.49 \textwidth]{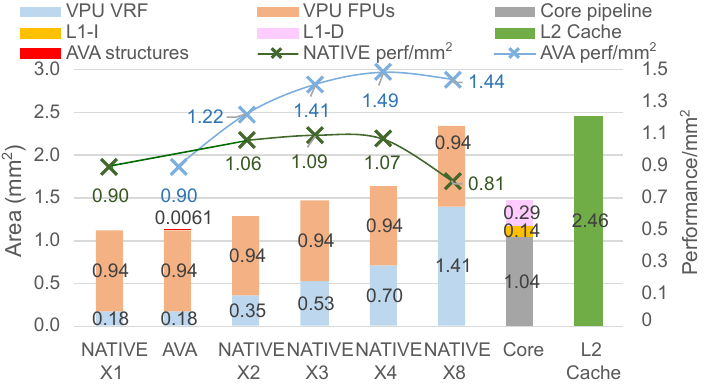}
\caption{Area results obtained from McPAT for 22nm technology node, and average performance/mm2 for each configuration.}
\label{figure4}
\end{figure}

To demonstrate the area efficiency of AVA, we modeled AVA and the five NATIVE configurations presented in Table 3 on the McPAT framework for 22nm technology. Figure~\ref{figure4} (left axis) shows the area results for all the VPU configurations presented in Table 3. We also include the area of the scalar core including L1I and L1D caches, and the 1MB L2 cache. AVA structures add a negligible 0.55\% area overhead to the VPU, while reducing the total VPU area by 53\% compared with the NATIVE X8 configuration.
 
To demonstrate the AVA performance/mm2 efficiency, we obtain the average performance for all the previous evaluated applications for NATIVE and AVA configurations, and divided between the area for the corresponding configuration. Results are shown in Figure~\ref{figure4} (right axis). Note that for AVA, the area is 1.126mm2 for all the configurations.

To demonstrate that AVA is energy efficient, we also obtain the Energy consumption for all the previous evaluated applications. Last column (left axis) of Figure 3 shows the energy consumption results. The application statistics introduced in the McPAT model corresponds to the gem5 outputs. Dynamic and leakage energy results are reported only for the main contributors: The L2 cache, the VRF and the FPUs.

The required AVA structures also are modeled, however it represents only 0.4\% of the overall VPU energy consumption for the AVA X1 configuration. Since the number of issued instructions is reduced as the MVL value is increased, the energy consumed by the required AVA structures is also reduced for larger MVL configurations. Then, we include the extra energy dissipation of AVA in the VRF Dynamic/Leakage bars for all the AVA configurations. 

Axpy (Figure 3-a4) and Particle-Filter (Figure 3-c4) shows a similar behavior since there are either no or few spill/swap operations. For both, as the MVL is increased, less total energy is consumed. Dynamic energy is constant since no spill/swap operations are added.  Since larger configurations improve performance, leakage energy is reduced for L2 and FPUs. However, the leakage for the VRF is different: NATIVE X2, X3, X4 and X8 configurations doubles the leakage in each configuration because they are implementing larger multi-ported VRF memories from 16KB up to 64KB. Then, both RG and AVA configurations consume less energy than the equivalent NATIVE configuration. For Axpy, when comparing with the NATIVE X1 configuration, AVA saves 37\% of the overall energy consumption by reconfiguring for long vectors.

Blackscholes (Figure 3-b4) and Swaptions (Figure 3-f4) generates an important number of spill/swap operations for the RG-LMUL8 and AVA X8, leading to extra energy dissipation which is wasted to support those operations. For Blackscholes, this leads to 13\% and 17\% more energy for RG and AVA respectively. 

Somier represents a memory bound application, where L2 leakage dominates the overall energy consumption. VRF leakage for NATIVE X8 also represent an important energy contributor. When comparing AVA X8 with NATIVE X8, it is clear the advantage of having an 8KB VRF where leakage contribution does not cause a big impact on the overall energy consumption.

Finally, LavaMD2 has interesting results. Energy consumption increases notably for RG-LMUL8 and AVA X8. This is because the application represents medium-size vectors, with MVL=48 being the optimal. Spill/swap operations are always executed with the MVL value. For RG-LMUL8 and AVA X8 configurations, spill/swap operations are executed with a MVL=128 although elements past VL=48 are not used, leading to a drastic energy consumption increase. However, when running the application, AVA will select the optimal configuration (AVA X3) avoiding to waste unnecessary energy.

As demonstrated with our results, AVA not only provides the capability to execute longer vectors and improve performance, but also saves energy. 

\section{Synthesis and place-and-route}

Finally, we also perform experiments with design automation tools to get accurate results for area and achievable frequency. Towards this goal, we added the required AVA support to an in-house VPU. We present synthesis and place-and-route results for AVA and NATIVE X8 configurations. To provide the 4R-2W VRF, we implemented the LVT technique~\cite{multiport} which provides multi-ported memories at the cost of replicating and banking dual-port memories.

We obtain the main physical metrics using Cadence tools, Genus for synthesis and Innovus for place-and-route. We selected the GLOBALFOUNDRIES 22FDX 8T technology libraries, and we implemented the VRF slices using the Synopsys High-Performance Dual-Port SRAM cell-based Register File Memory Compiler (R2PH). The target frequency was 1GHz.

\begin{scriptsize}
\begin{table}[h]
\setlength{\tabcolsep}{0.8\tabcolsep}
    \caption{Post-place-and-route results.}
  \label{table:table3}
  \centering
  \begin{tabular}{ | c | c | c | c | c |}
    \hline
  			 & \textbf{WNS (ns)} & \textbf{Power  (mW)} & \textbf{Area (mm2)} & \textbf{Density} \\
    \hline
    \hline
    NATIVE X8  		& -0.244 &   2290 	& 3.90  	& 61.0\% 	\\
      	-VRF macros  	&      	&   388 	& 1.252 	& 			 \\
    \hline
     AVA  		& +0.119 &   1732 	& 1.98  	& 61.8\% 	\\
	-AVA strctures  &      	&   5.266 	& 0.0042 	& 			 	\\
	-VRF macros  	&      	&   184 	& 0.257 	& 			 	\\
    \hline
  \end{tabular}
\end{table}
\end{scriptsize}

Post-place-and-route results for the typical corner (TT 0.8V 25Cº) are summarized in Table 5, and the obtained layouts are shown in Figure 5, for both configurations. Regarding area results, for the AVA configuration, the required AVA hardware structures incur a negligible 0.21\% area overhead. On the other hand, the total chip area is reduced by 50.7\% compared with the NATIVE X8 configuration, validating the McPAT results.

Regarding the timing performance, target constraints are met only for AVA with a positive slack of 0.119ns. However, for NATIVE X8 there is a negative slack of -0.244ns, due to the critical paths stemming from the longer wires between the SRAMs and the lane logic. Based in our synthesis and place-and-route experiments, we can confirm that the small size required for AVA helps to achieve higher working frequencies due to a higher robustness against different physical floorplanning options.

\begin{figure}[h!]
\centering
\includegraphics[width=0.5\textwidth]{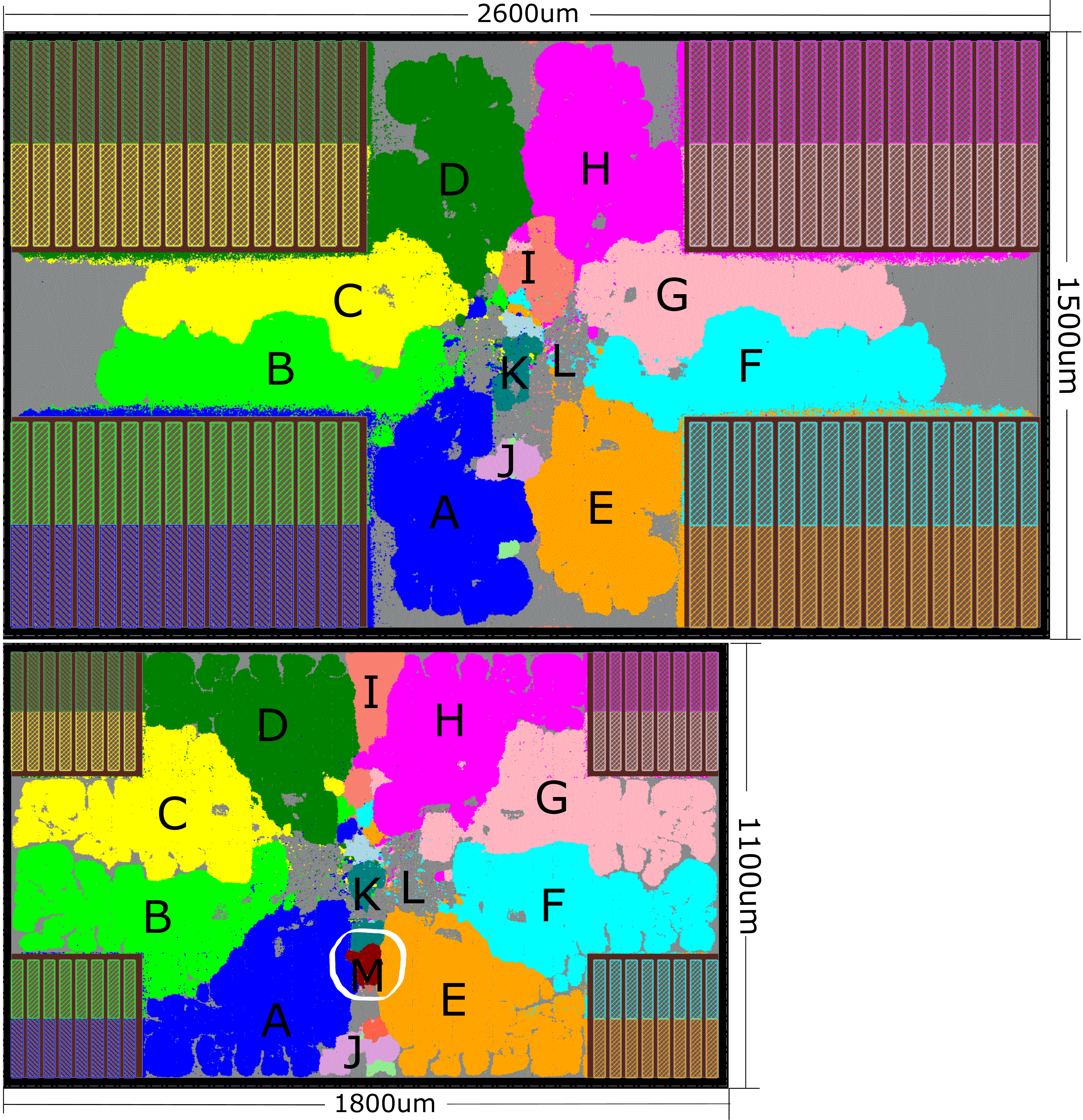}
\caption{PnR results for 22 nm technology of two instances of the VPU with eight lanes, highlighting main internal blocks: A) lane 1; B) lane 2; C) lane 3; D) lane 4; E) lane 5; F) lane 6; G) lane 7; H) lane 8; I) Vector Memory Unit; J) ROB; K) Instruction queue; L) Remaining modules such as Memory queue, Rename Unit and Ring Lane Interconnection; M) AVA structures. NATIVE X8 design is on the top and AVA on the bottom. Circled area marks the added AVA structures. VRF memory macros can be identified on the corners.}
\label{figure5}
\end{figure}

\section{Related Work}

AVA partially leverages different computer architecture techniques that were developed for out-of-order cores, VLIW processors and GPUs. While the concepts might be familiar at high level, we adapt and substantially tailor these techniques for VPs to propose the novel adaptable VRF design. Next lines briefly describe the related work.

Different alternatives to exploit efficient use of physical registers was widely studied. González et al. ~\cite{virtualreg,Monreal} proposed a dynamic register renaming approach where the key idea is to delay the allocation of the physical registers until write-back. To this end, a technique termed as Virtual-Physical Registers was proposed. Virtual-Physical Registers are not related to any storage location; they are merely tags to keep track of the dependencies and are therefore not related to AVA. Although AVA proposes a two-level renaming scheme, unlike the Virtual-Physical Registers concept, our VVRs are assigned at renaming time, while physical registers are assigned at issue time, and combined with the RAC counters, exploiting the use of the vector registers as soon as they can be reused. 

Based on the fact that a physical register can be reused when it is guaranteed that the value in it can never be used by any later instruction, several studies~\cite{Akkary,Moudgill,Smith}    associated a counter with each physical register, to keep track of the pending read operations. In these techniques, a physical register is freed whenever the associated counter is zero. Such aggressive register reclamation schemes enable physical register usage to closely match the true lifetime of registers. AVA exploits the concept of aggressive register reclamation to free a physical register that will not be longer used. Additionally, AVA extend the use of the associated counters to decide the best option to perform swaps between Physical and VVRs. 

The idea of using memory to provide a backing store to the register file has been has also been widely studied for out-of-order cores~\cite{fake1000} , VLIW processors~\cite{twolevelrf}, and GPUs~\cite{rf-GPUs,Regless}. In this work, we apply it to VPs as a key mechanism to offer a variety of MVL configurations. Additionally, we have unified the idea of a two-level VRF with the concept of VVRs and physical registers, which in combination with the Swap Mechanism presents a balanced design which is able to efficiently handle different DLP patters.

\section{Conclusions}

This paper introduces AVA, an Adaptable Vector Architecture with the ability to reconfigure the MVL, unlocking the benefits of having a longer vector microarchitecture when abundant DLP is present. Our results demonstrate that by having a modest VPU designed for short vectors, plus our novel scheduling mechanism, it is possible to obtain a very competitive performance when comparing AVA with the equivalent native long vector configurations. As a first approximation, we obtain area and energy metrics from McPAT, demonstrating that AVA can save around 53\% of the total VPU area compared with a native configuration for long vectors. Additionally, we demonstrate that supporting long vectors not only improve performance, but also leads to energy savings for several workloads. Finally, we implemented AVA at RTL level, synthesized and place-and-routed in 22nm technology, demonstrating that AVA not only provides an area-efficient design, but also allows higher frequencies.


\section*{Acknowledgment}

Research reported in this publication is partially supported by CONACyT Mexico under Grant No. 472106, the Spanish State Research Agency - Ministry of Science and Innovation (contract PID2019-107255GB), and the DRAC Project under Grant 001-P-001723.

\bibliographystyle{IEEEtranS}
\bibliography{refs}

\end{document}